\documentclass[10pt, journal, letterpaper]{IEEEtran}
\usepackage[ruled, vlined]{algorithm2e}
\usepackage{algpseudocode}
\usepackage{amsmath}
\usepackage{amssymb}
\usepackage{array}
\usepackage{arydshln}
\usepackage{blkarray, bigstrut}
\usepackage{bm}
\usepackage{booktabs}
\usepackage{braket}
\usepackage{cite}
\usepackage{cuted}
\usepackage{enumitem}
\usepackage{graphicx}
\usepackage[mathscr]{euscript}
\usepackage{mathtools}
\usepackage{multirow}
\usepackage{multicol}
\usepackage{framed} 
\usepackage[framed]{ntheorem}
\usepackage{nicefrac}
\usepackage{pgfplots}
\usepackage{qtree}
\usepackage{adjustbox}
\usepackage{rotating}

\pgfplotsset{compat = 1.13}
\usepackage{stfloats}
\usepackage{subfig}
\usepackage{url}
\usepackage{tikz}
\usepackage[framemethod=TikZ]{mdframed}
\usetikzlibrary{patterns}
\usetikzlibrary{spy}
\usetikzlibrary{shapes, backgrounds,calc, snakes}
\usetikzlibrary{decorations.pathreplacing}
\usepgfplotslibrary{fillbetween}
\usetikzlibrary{matrix}
\usepackage{fancybox}

\tikzstyle{vertex} = [circle, draw, inner sep = 0pt, minimum size = 10pt]

\newframedtheorem{frm-prop}{Property}

\definecolor{bblue}{rgb}{0.12392, 0.0490, 0.9588}
\definecolor{sskyblue}{rgb}{0.1529, 0.5882, 0.9216}
\definecolor{ggreen}{rgb}{0.5020, 0.7961, 0.3451}
\definecolor{yyellow}{rgb}{0.9765, 0.9804, 0.0784}

\definecolor{color0}{HTML}{FF0147}
\definecolor{color1}{HTML}{F400DC}
\definecolor{color2}{HTML}{BA0DFF}
\definecolor{color3}{HTML}{5700E8}
\definecolor{color4}{HTML}{0B03FF}
\definecolor{color5}{HTML}{0957F4}
\definecolor{color6}{HTML}{03B3FF}
\definecolor{color7}{HTML}{08E8DA}
\definecolor{color8}{HTML}{07FF8E}
\definecolor{color9}{HTML}{51FF0A}

\definecolor{p1}{rgb}{1, 0.0667, 0}
\definecolor{p2}{rgb}{1, 0.24, 0}
\definecolor{p3}{rgb}{1, 0.349, 0}
\definecolor{p4}{rgb}{1, 0.490, 0}
\definecolor{p5}{rgb}{1, 0.631, 0}
\definecolor{p6}{rgb}{1, 0.792, 0}
\definecolor{p7}{rgb}{1, 0.933, 0}
\definecolor{p8}{rgb}{1, 1, 0}
\definecolor{p9}{rgb}{1, 1, 0.5}
\definecolor{p10}{rgb}{1, 1, 0.8}

\begin{document}

\title{System Level Simulation of Scheduling Schemes for C-V2X Mode-3}

\author
{
	\IEEEauthorblockN{Luis F.~Abanto-Leon\IEEEauthorrefmark{2}, Arie Koppelaar\IEEEauthorrefmark{3}, Chetan B. Math\IEEEauthorrefmark{4}, Sonia Heemstra de Groot\IEEEauthorrefmark{8}} \\
	\IEEEauthorblockA{\IEEEauthorrefmark{2} \IEEEauthorrefmark{4} \IEEEauthorrefmark{8} Eindhoven University of Technology}
	\IEEEauthorblockA{\IEEEauthorrefmark{3}NXP Semiconductors} \\ 
	\IEEEauthorrefmark{2}l.f.abanto@tue.nl,
	\IEEEauthorrefmark{3}arie.koppelaar@nxp.com,
	\IEEEauthorrefmark{4}c.belagal.math@tue.nl,
	\IEEEauthorrefmark{8}sheemstradegroot@tue.nl,
}

\maketitle

\begin{abstract}
	The 3rd Generation Partnership Project (3GPP) introduced Cellular Vehicle-to-Everything (C-V2X) as a novel technology for enabling sidelink vehicular communications. While a distributed scheduling scheme (i.e., C-V2X \textit{mode-4}) has been standardized by 3GPP in order to support out-of-coverage scenarios, the design of centralized infrastructure-based schemes (i.e., C-V2X \textit{mode-3}) is open to implementation. In this paper, we propose two C-V2X \textit{mode-3} schemes based on bipartite graph matching (BGM), called \texttt{BGM - Pow} and \texttt{BGM - Dist}. Specifically, \texttt{BGM - Pow} allocates subchannels based on the minimization of the overall power perceived by the vehicles whereas \texttt{BGM - Dist} is based on the maximization of the subchannels reusage distance. Through simulations we show that the proposed centralized schemes outperform C-V2X \textit{mode-4} as the subchannels can be assigned more efficiently with reduced interference.
\end{abstract}

\begin{IEEEkeywords}
	subchannel scheduling, vehicular networks, mode-3, mode-4, sidelink, SPS, C-V2X, LTE-V2X.
\end{IEEEkeywords}

\IEEEpeerreviewmaketitle

\section{Introduction}
Cellular Vehicle-to-Everything (C-V2X) communications is one of the novel paradigms introduced by the 3rd Generation Partnership Project (3GPP) \cite{3gpp.36.213} in Release 14. C-V2X communications is envisaged as a dependable technology capable of meeting stringent latency and reliability requirements in dynamic vehicular scenarios. Two operation modalities have been defined within C-V2X, namely \textit{mode-3} and \textit{mode-4} \cite{3gpp.36.213}. The former aims centralized systems, and therefore relies on the availability of cellular infrastructure, such as eNodeBs. Specially, eNodeBs can optimize any utility criteria for allocating sidelink subchannels to vehicles. For instance, \cite{abanto2017:graph-resource-allocation-conflict-avoidance-v2v-broadcast-communications} describes a scheme in which the sum-rate capacity is maximized based on the subchannels signal-to-interference-plus-noise ratio (SINR) that vehicles report to eNodeBs. Contrastingly, \cite{abanto2018:subchannel-allocation-v2v-broadcast-communications-mode3} relies on the geographical clustering of vehicles, where intra- and inter-cluster constraints are defined with additional differentiated QoS requirements per vehicle. Once an eNodeB has computed a suitable allocation of the subchannels, vehicles are notified via downlink---upon which direct vehicle-to-vehicle (V2V) sidelink communications takes place over the assigned subchannels. Contrastingly, C-V2X \textit{mode-4} operates autonomously and distributedly without the necessity of a central orchestrator. Thus, vehicles monitor the power levels throughout all the subchannels and then select a suitable one for their own utilization.

\noindent \textbf{\textit{Dichotomy between centralized and distributed schemes:}} In C-V2X \textit{mode-3}, the sidelink subchannels can be allocated more efficiently due to the humongous amount of knowledge that is available at the eNodeBs. Conversely, a noticeable drawback of C-V2X \textit{mode-4} is the restricted local knowledge of each vehicle, thus preventing them from selecting an optimal subchannel. Moreover---due to incoordination---several vehicles may compete over the same subset of subchannels, thus leading to severe packet reception ratio (PRR) degradation. In order to diminish the occurrences of allocation conflicts, 3GPP standardized a semi-persistent scheduling (SPS) scheme. In particular, \emph{(i)} each vehicle senses the power on all subchannels to understand their utilization (i.e. busyness) and \emph{(ii)} then selects a subchannel, which may be unoccupied or experiences low received power. The allocated subchannel is utilized on a quasi-steady basis by the vehicle until rescheduling is required. Due to short-term invariability of the allocation, any vehicle seeking rescheduling is capable of acquiring a degree of understanding on the subchannels utilization by performing sensing. Thus, through this mechanism, vehicles can improve the likelihood of its own transmitted messages being received reliably by selecting suitable (but not optimal) subchannels with low interference. 


\noindent \textbf{\textit{Motivation:}} While C-V2X \textit{mode-4} has been clearly defined, there is a lack of centralized scheduling schemes for C-V2X \textit{mode-3} communications. In this paper, we have devised two approaches inspired by bipartite graph matching (BGM) and compare them against C-V2X \textit{mode-4} proposed by 3GPP \cite{3gpp.36.213}. 

The paper is structured as follows. In Section II, we describe the system model. In Section III, the two proposed C-V2X \textit{mode-3} scheduling approaches are described. In Section IV, the 3GPP SPS scheduling scheme for C-V2X \textit{mode-4} communications is briefly revisited. Section V is devoted to simulation results. Finally, Section VI summarizes our conclusions.

\section{System Model}
A 10 MHz intelligent transportation systems (ITS) channel for exclusive support of sidelink communications is assumed. The whole channel is divided into several time-frequency resource partitions---hereinafter called subchannels. Each subchannel has dimensions of 1 ms in time (i.e. equivalent to one LTE subframe length) and $ N $ resource blocks (RBs) in frequency. A subchannel is assumed to be capable of carrying a cooperative awareness message (CAM). In this work, we assume a nominal message rate of $\Delta_{CAM} = 10$ Hz. Therefore, the maximum amount of time divisions is 100. In a similar manner as subchannels in C-V2X \textit{mode-4} are reserved on a semi-persistent basis---in the centralized schemes presented herein---the same assumption prevails in order to make a fair comparison between the approaches. Thus, subchannels are utilized by vehicles for $T_{SPS}$ seconds ($T_{SPS}$ is random as specified by 3GPP). During this period, a vehicle will periodically broadcast on the assigned subchannel, and upon termination a new reservation will be required. In addition, at any time instance there is a number of vehicles in the system that require re-scheduling (when $T_{SPS}$ countdown has reached zero). The eNodeB will be notified of such a requirement and a new allocation of resources will be attempted using the schemes discussed in Section III.

\section{Proposed Scheduling for C-V2X Mode-3}
In this section, we briefly describe two schedulings approaches that we propose. 

\noindent{\textbf{\texttt{BGM-Pow}} (\textit{Minimization of overall received power}):} In this approach the objective is to minimize the overall received power of the assigned subchannels. Thus, vehicles requiring re-scheduling transmit via uplink the conditions (i.e., sensed power) they experience on the monitored subchannels. Based on this information, a distribution of resources that minimizes the total received power will be computed. The intuition behind this approach is that subchannels with low received power may be more suitable as they might be unoccupied or experience negligible interference. In this case, the vertices of the graph are either vehicles or subchannels, and the edge weights are represented by the RSSI values of the subchannels.

\noindent{\textbf{\texttt{BGM-Dist}} (\textit{Maximization of subchannel reusage range}):} The objective of this approach is to attain maximal reusage distance for every subchannel in the system. Intuitively, this will help in mitigating co-channel interference. In this case, the edge weights are determined by the minimum distance at which other vehicles are reusing a particular subchannel.

\section{3GPP-based Scheduling for C-V2X Mode-4}
In the following, we briefly describe the 3GPP C-V2X \textit{mode-4} scheduling scheme. For a more detailed explanation, the reader is referred to \cite{3gpp.36.213} \cite{3gpp.36.331}. It consists of the following stages.

\begin{center}
	\begin{table}[!t]
		\centering
		\scriptsize
		\caption {Simulation parameters}
		\label{t1}
		\begin{tabular}{lccc}
			\toprule
			\multicolumn{1}{c}{\textbf{Description}} & \multicolumn{1}{c}{\textbf{Symbol}} & \multicolumn{1}{c}{\textbf{Value}} & \multicolumn{1}{c}{\textbf{Units}}\\
			\midrule
			Number of RBs per subchannel (per subframe) & $ N $ & 30 & - \\
			Number of sub-bands & $F$ & 3 & - \\
			Number of subchannels & - & 300 & -\\
			CAM message rate & $\Delta_{CAM}$ & 10 & Hz \\
			CAM size & $M_{CAM}$ & 190 & bytes\\
			MCS & - & 7 & - \\
			Transmit power per CAM & -& 23 & dBm \\
			SINR threshold & $\gamma_{T}$ & 3.98 & dB \\
			Scheduling period \cite{3gpp.36.331} & $T_{SPS}$ & 0.5-1.5 & s \\
			Antenna gain & $G_t, G_r$ & 3 & dB \\
			Shadowing standard deviation & $\mathcal{X}_{\sigma}$ & 7 & dB \\
			Shadowing correlation distance & - & 10 & m \\
			\bottomrule
		\end{tabular}
	\end{table}
	\vspace{-0.75cm}
\end{center}

\noindent{\textbf{\textit{Power sensing}:}} The vehicles continuously monitor the received power on each subchannel except on those where monitoring was not possible, e.g. due to half-duplex limitation.

\noindent{\textbf{\textit{Subchannels categorization}:}} Some subchannels will be excluded from selection based on the average PSSCH-RSRP. This is an iterative stage where a threshold is gradually increased in order to admit more subchannels with incremented interference until there are at least 20\% of candidate subchannels for selection. 

\noindent{\textbf{\textit{Subchannel selection}:}} Among the pre-selected candidates in the previous stage, a subchannel is randomly chosen.

\section{Simulations}
In this section, we compare the standardized 3GPP scheduling method against the two proposed approaches. We assess a freeway scenario with 600 vehicles over 40 seconds of simulation. The simulations have been performed 5 times with different seeds in order to evaluate deviations. The relevant parameters for the experiments are shown in Table \ref{t1}.

We observe in Fig. \ref{f1} the PRR curves for the mentioned approaches. Furthermore in shaded colors, we depict the deviations which are originated due to performance variations when using different seeds. These representations can be regarded as intervals with 99.999\% of confidence. We can observe that the scheduling scheme based on distance maximization outperforms C-V2X \textit{mode-4} across all the distances. However, the approach based on minimization of the sum of received powers can only outperform C-V2X \textit{mode-4} in the near field---which is a critical region. In the far-field (150 m an beyond) its performance is comparable to C-V2X \textit{mode-4}.

\begin{figure}[!t]
	\centering
	\begin{tikzpicture}[line cap=round,line join=round,x=2cm,y=2cm,spy using outlines={rectangle,lens={scale=3}, size=8cm, connect spies}]
	\begin{axis}[
	xmin = 50,
	xmax = 300,
	ymin = 0.48,
	width = 9.0cm,
	height = 5.5cm,
	xlabel={Distance [meters]},
	x label style={align=center, font=\footnotesize,},
	ylabel = {Packet Reception Ratio (PRR)},
	y label style={at={(-0.08,0.5)}, text width = 3.5cm, align=center, font=\footnotesize,},
	ytick = {0.5, 0.6, 0.7, 0.8, 0.9, 1.0},
	yticklabels = {0.5, 0.6, 0.7, 0.8, 0.9, 1.0},
	legend columns = 1,
	legend style={at={(0.008,0.01)},anchor=south west, font=\fontsize{6}{7}\selectfont, text width=3.1cm,text height=0.05cm,text depth=.ex, fill = none, align = left},
	]
	
	\addplot[color=black, mark = diamond*, mark options = {scale = 1.5, fill = cyan}, line width = 0.5pt] coordinates  
	{
		(50, 0.9949)
		(100, 0.9761)
		(150, 0.8964)
		(200, 0.7690)
		(250, 0.6272)
		(300, 0.4902)
	}; \addlegendentry{\texttt{Proposed} $|$ \texttt{BGM-Pow}}

	\addplot[color=black, mark = square*, mark options = {fill = orange, solid}, line width = 0.5pt] coordinates
	{
		(50, 0.9847)
		(100, 0.9574)
		(150, 0.8926)
		(200, 0.7996)
		(250, 0.6824)
		(300, 0.5514)
	}; \addlegendentry{\texttt{Proposed} $|$ \texttt{BGM-Dist}}

	\addplot[color=black, mark options = {fill = yyellow}, line width = 0.5pt, style = densely dotted] coordinates 
	{
		(50, 0.9814)
		(100, 0.9516)
		(150, 0.8767)
		(200, 0.7662)
		(250, 0.6365)
		(300, 0.5048)
	}; \addlegendentry{\texttt{Standard} $|$ \texttt{3GPP Mode-4}}

	\addplot[name path = Al, color=cyan!40, line width = 0pt] coordinates  
	{
		(50, 0.9937)
		(100, 0.9749)
		(150, 0.8933)
		(200, 0.7625)
		(250, 0.6196)
		(300, 0.4821)
	};

	\addplot[name path = Au, color=cyan!40, line width = 0pt] coordinates  
	{
		(50, 0.9961)
		(100, 0.9773)
		(150, 0.8994)
		(200, 0.7756)
		(250, 0.6348)
		(300, 0.4982)
	};

	\addplot[cyan!40, fill opacity=0.4] fill between[of = Al and Au];

	\addplot[name path = Bl, color=orange!40, line width = 0pt] coordinates  
	{
		(50, 0.9826)
		(100, 0.9548)
		(150, 0.8879)
		(200, 0.7901)
		(250, 0.6767)
		(300, 0.5460)
	};
	
	\addplot[name path = Bu, color=orange!40, line width = 0pt] coordinates  
	{
		(50, 0.9869)
		(100, 0.9599)
		(150, 0.8972)
		(200, 0.8090)
		(250, 0.6880)
		(300, 0.5568)
	};
	
	\addplot[orange!40, fill opacity=0.4] fill between[of = Bl and Bu];
	
	\addplot[name path = Cl, color=black!20, line width = 0pt] coordinates  
	{
		(50, 0.9781)
		(100, 0.9489)
		(150, 0.8711)
		(200, 0.7558)
		(250, 0.6272)
		(300, 0.4958)
	};

	\addplot[name path = Cu, color=black!20, line width = 0pt] coordinates  
	{
		(50, 0.9848)
		(100, 0.9544)
		(150, 0.8823)
		(200, 0.7766)
		(250, 0.6457)
		(300, 0.5137)
	};
	
	\addplot[black!20, fill opacity=0.4] fill between[of = Cl and Cu];
	
	\end{axis}
	
	\end{tikzpicture}
	\caption{PRR performance for various distances}
	\label{f1}
	\vspace{-0.2cm}
\end{figure}

\section{Conclusion}
We have proposed two centralized scheduling schemes for C-V2X \textit{mode-3}. We conclude that only the distance-based approach performs better than the distributed mode for all distance ranges. For further study, we aim at combining both RSSI values and distance information to improve the performance of the centralized approach.

\bibliographystyle{IEEEtran}
\bibliography{ref}
	


\end{document}